\documentclass[11pt,a4paper]{article}

\RequirePackage[l2tabu, orthodox]{nag}

\usepackage{jheppub}
\usepackage{amsthm,amssymb,amsmath,epic,float}
\usepackage{rotating,epsfig,indentfirst,array,varioref}
\usepackage{appendix,tikz,mathtools}
\usepackage{setspace}
\usepackage{tikz}
\usetikzlibrary{decorations.pathreplacing}
\usetikzlibrary{calc}
\usepackage{enumitem}
\usepackage{hyphenat}
\usetikzlibrary{positioning}
\usepackage{graphicx}  
\usepackage{adjustbox}

\usepackage{arydshln}
\usepackage{cancel}

\usepackage{subfig}
\usetikzlibrary{intersections}
\pgfdeclarelayer{background}
\pgfsetlayers{background,main}
\newcommand{\bref}[1]{\textbf{\ref{#1}}}
\usepackage{tikz,pgf}
\usetikzlibrary{shapes}
\usetikzlibrary{calc}
\usetikzlibrary{decorations.pathmorphing}
\usetikzlibrary{decorations.pathreplacing,shapes.misc}
\usetikzlibrary{positioning}
\usetikzlibrary{arrows}
\usetikzlibrary{decorations.markings}
\usetikzlibrary{shadings}
\usetikzlibrary{intersections}
\def\be{\begin{equation}}
\def\ee{\end{equation}}

\def\be{\begin{equation}}
\def\en{\end{equation}}
\def\ber{\begin{eqnarray}}
\def\enr{\end{eqnarray}}

\newcommand{\ket}[1]{\left| #1 \right\rangle}
\newcommand{\bra}[1]{\left\langle #1 \right|}
\newcommand{\pd}{\partial}

\newcommand{\br}[1]{{\overline{#1}}}

\def\<{\left(}
\def\>{\right)}

\usepackage{jheppub}
\makeatletter
\def\@fpheader{\vspace{-.1cm}}
\makeatother

\title{\boldmath  Torus one-point correlation numbers in minimal Liouville gravity }

\author[a,b]{A.~Artemev,}
\author[c]{V.~Belavin}

\affiliation[a]{Landau Institute for Theoretical Physics, 142432 Chernogolovka, Russia}
\affiliation[b]{Skolkovo Institute of Science and Technology, 121205, Moscow, Russia}
\affiliation[c]{Physics Department, Ariel University, Ariel 40700, Israel}

\emailAdd{artemev.aa@phystech.edu}
\emailAdd{vladimirbe@ariel.ac.il}

\abstract{
We present a method for the first principles calculation of tachyon one-point amplitudes in $(2,2p+1)$ minimal Liouville gravity defined on a torus. The method is based on the higher equations of motion in the Liouville CFT. These equations were earlier successfully applied for analytic calculations of the amplitudes in the spherical topology. We show that this approach allows to reduce the moduli integrals entering the definition of the torus amplitudes to certain boundary contributions, which can be calculated explicitly. The results agree with the calculations performed in the matrix models approach.}


\begin{document} 
\maketitle
\flushbottom

\section{Introduction}
\label{sec:intro}

Minimal Liouville gravity (MLG) is a model of non-critical string theory~\cite{Polyakov1}. 
In a series of papers~\cite{Belavin:2013nba,Belavin:2014cua, Belavin:2014hsa,Belavin:2014xya,Belavin:2015ffa} for some correlators in genus zero it was shown that MLG results can also be obtained from 
Matrix Models~\cite{Douglas:1989dd} (in the double-scaling limit) provided appropriate identifications between coupling constants~\cite{MSS,Belavin:2008kv}. In the continuous approach an essential progress has been achieved thanks to discovery of the so-called higher equations of motion (HEM) in the Liouville conformal field theory~\cite{HEM}. 
However it was not clear whether it is possible to develop a method based on HEM for considering higher genus topology. Even though some numerical checks were performed~\cite{Belavin:2010pj,Belavin:2010sr},  which gave evidence that the correspondence between MLG and Matrix Models persists at higher genus, the proof was not possible because the analytic calculation of the moduli integral in the torus topology was not available.  The purpose of this paper is to describe a method for computing the amplitudes in MLG for a basic class of  physical observables,  which are correlation functions of tachyon vertex operators, in the case of torus topology.

The plan of the paper is as follows. In Section~\ref{sec:2} we recall some  facts about continuous formulation of MLG, which are relevant for our consideration.
In Section~\ref{sec:3} we consider the particular examples of  the torus one-point  correlation numbers.
Section~\ref{sec:concl} is a conclusion in which we discuss in particular some open questions.

\section{Preliminaries}
\label{sec:2}
\subsection{Generalities of minimal models and Lioville field theory}
\paragraph{Liouville CFT. }Liouville conformal field theory is a 2-dimensional CFT, which has the action of the form
\begin{equation}
    A_L = \int d^2x\,\sqrt{\hat{g}}\left(\frac{1}{4\pi}\hat{g}^{ab} \pd_a \phi \pd_b \phi + \mu e^{2b\phi} + \frac{Q}{4\pi} \hat{R} \phi \right)\;.
\end{equation}
Here $\hat{g}$ is a reference metric with scalar curvature $\hat{R}$, introduced to have the theory in a covariant form. The parameter $Q = b + b^{-1}$; the Virasoro central charge of the theory is then given by $c_L = 1 + 6 Q^2$. $\mu$ is the cosmological constant parameter; the dependence of the correlation functions on $\mu$ is fixed, e.g., by noticing that $\mu$ can be set to one by shifting $\phi \to \phi - \frac{1}{2b} \log \mu$.

The normal ordered exponential operators $V_a = :e^{2a \phi}:$ are primary fields of the model of conformal dimension $\Delta^\text{L}_a = a (Q-a)$. Operators $V_a$ and $V_{Q-a}$ have the same conformal dimension; they are identified up to a factor $R_L(a)$, which is called the ``reflection coefficient''.  

The so-called degenerate fields  $V_{m,n} \equiv V_{a_{m,n}}$ in the Liouville theory arise for the special values of the dimension parameter
\begin{equation}
a_{m,n} = - b^{-1} \frac{(m-1)}{2} - b \frac{(n-1)}{2} \;.\label{lioprim}
\end{equation}
These are the primary fields (Virasoro highest vectors) with a descendant at level $mn$ in the corresponding highest weight representation, which itself is a highest vector. In other words, the corresponding Verma module has a submodule, in CFT such submodules are usually decoupled by putting its highest vector to zero. Decoupling conditions can be written as $D^{(L)}_{m,n} V_{m,n} = 0$, where $D^{(L)}_{m,n}$ is a polynomial of Virasoro generators $L_{-k}$ of degree $mn$: $D^{(L)}_{m,n} = L_{-1}^{mn} + \dots\,$. This condition leads to the fact that only finite number of conformal families, the so called fusion channels, contribute to the OPE with a degenerate field. 

An expression for the sphere three-point functions (or structure constants) of Liouville theory $C_L(a_1, a_2, a_3)$ was first proposed by two groups of authors \cite{dornotto1994} and \cite{LFT}. It can be motivated e.g. by deriving recursion relations following from conformal bootstrap/crossing symmetry equations for four-point correlator with degenerate field $V_{1,2}$. A relevant solution of these relations is called Dorn-Otto-Zamolodchikov-Zamolodchikov (DOZZ) three-point function
%. The formula is
\begin{equation}
C_L(a_1, a_2, a_3) = (\pi \mu \gamma(b^2) b^{2-2b^2})^{(Q-a)/b} \frac{\Upsilon_b(b) \Upsilon_b(2a_1)\Upsilon_b(2a_2)\Upsilon_b(2a_3)}{\Upsilon_b(a-Q) \Upsilon_b(a-2a_1) \Upsilon_b(a-2a_2)\Upsilon_b(a-2a_3)}\;, \label{liouv3pf}
\end{equation}
where $\Upsilon_b(x)$ is a certain special function,
which obey two important properties: first, the shift relations 
\begin{align}
    & \Upsilon_b (x+b) = \gamma(bx) b^{1-2bx} \Upsilon_b(x)\;,\\
    & \Upsilon_b (x + b^{-1}) = \gamma(x/b) b^{2b/x-1} \Upsilon_b(x)\;,
\end{align}
and second is that this function has zeroes for $x= - m b - \frac{n}{b}$ and $x = Q + \frac{m}{b} + n b$ for  non-negative integers $m,n$. 

The OPE $V_{a_1}(x) V_{a_2}(0)$ in Liouville CFT is most simply written when $a_1, a_2$ lie in the so-called ``basic domain'' defined by
\begin{equation}
    \left|\frac{Q}{2} - \text{Re }a_1\right| + \left|\frac{Q}{2} - \text{Re }a_2\right| < \frac{Q}{2} \label{basicdom}\;.
\end{equation}
In this case one can write
\begin{equation}
    V_{a_1}(x) V_{a_2}(0) = \int \limits_{-\infty}^{\infty} \frac{dP}{4\pi} C_L(a_1, a_2, \frac{Q}{2}- i P) (x \br{x})^{\Delta^\text{L}_{Q/2 + i P} - \Delta^\text{L}_{a_1} - \Delta^\text{L}_{a_2}} [V_{Q/2+iP} (0)]\;,
\end{equation}
where $[\dots]$ stands for the contribution of the Virasoro representation associated with the corresponding primary field.
The operators in the Liouville theory are parametrized by complex $a$, but OPE is given in terms of contour integral and fields that appear here are parametrized by one real number $P$. This is because only operators with $a = \frac{Q}{2} + i P,\,P \in \mathbb{R}$, correspond to normalizable, or ``physical'', states of the theory. This is why in the OPE the sum goes over these states only. 

When parameters of the correlators are not in this domain, some poles of the structure constants may cross the integration contour. In order to keep the analyticity in the parameters in such a case one should either deform the contour of integration over $P$ or (equivalently) keep the contour unchanged but explicitly add the contributions from these poles which are referred to as ``discrete terms''. In fact, when one of the fused fields is degenerate, only a finite number of these terms contribute to the OPE (the ``continuous'' part becomes zero), as one could expect.

\paragraph{Minimal model. }Minimal models (MM) are CFTs in which only finite number of conformal families of operators exist. All primary fields have to be degenerate for consistency, so that after the fusion of two of them we would also have only a finite number of contributions. For the special values of the central charge defined by
\begin{equation}\label{C_M}
c_M = 1 - 6(\beta^{-1}-\beta)^2,\quad \beta = \sqrt{\frac{r}{r'}},\quad r<r' - \text{ coprime numbers }
\end{equation}
in the lattice of degenerate fields $\Phi_{m,n}$  there is a subset called the Kac table that consists of fields $\Phi_{m,n}$ with $0<m<r,\, 0<n<r'$, on which the OPE is closed.  There is also Kac symmetry identification $\Phi_{m,n} = \Phi_{r-m, r'-n}$, which is similar to the reflection property in the Liouville CFT. A theory that contains only conformal families from such a Kac table is called a minimal model $\mathcal{M}_{r,r'}$. The conformal dimensions of the primary fields are parametrized as $\Delta_\alpha^\text{M} = \alpha(\alpha-\beta^{-1}+\beta)$ with
\begin{equation}
 \alpha_{m,n} = \frac{\beta^{-1}-\beta}{2} + \frac{n\beta - m \beta^{-1}}{2}\;.
\end{equation}
The ``kinematical'' properties of the minimal model strongly remind of those of the Liouville theory for imaginary $b$. For example, the decoupling conditions defining the fusion rules have the form $D^{(M)}_{m,n} \Phi_{m,n} = 0$, where operator $D^{(M)}$ can be obtained from $D^{(L)}$ by replacing $b^2 \to -\beta^2$; the recursion relations allowing to obtain the structure constants are also given by analytic continuation $b \to -i \beta, a_i \to i\alpha_i$. 

Relevant for the minimal model solutions to the recursion relations above are not analytic continuations of DOZZ structure constants. In fact, they are given by \cite{Zamolodchikov:2005fy}
\begin{equation}
C_M (\alpha_1, \alpha_2, \alpha_3) = \frac{\mathcal{A}\Upsilon_{\beta} (\alpha - 2 \alpha_3 + \beta) \Upsilon_{\beta} (\alpha - 2 \alpha_1 + \beta) \Upsilon_{\beta} (\alpha - 2 \alpha_2 + \beta) \Upsilon_{\beta} (\alpha + 2\beta - \beta^{-1})}{\prod \limits_{i=1}^3 (\Upsilon_\beta(2 \alpha_i + \beta) \Upsilon_\beta(2\alpha_i +2 \beta - \beta^{-1}))^{1/2}} \;,\label{min3pf}
\end{equation}
where
\begin{equation}
    \alpha \equiv \alpha_1 + \alpha_2 + \alpha_3\;,\quad\mathcal{A} \equiv \frac{\beta^{\beta^{-2} - \beta^2 - 1} [\gamma(\beta^2) \gamma(\beta^{-2}-1)]^{1/2}}{\Upsilon_\beta(\beta)}\;.
\end{equation}
When specialized to the degenerate values of $b$ and $\alpha$, (\ref{min3pf}) coincides with the minimal model structure constants whenever they are non-zero, but it does not necessarily conform with the fusion rules (i.e., it can give a non-zero answer even if  the sphere three-point function $\langle \Phi_{\alpha_1} \Phi_{\alpha_2} \Phi_{\alpha_3} \rangle$ should be zero; most explicit example is that $C_M(\alpha_1, \alpha_2,0)$ can be non-zero even for $\Delta^{\text{M}}_{\alpha_1} \neq \Delta^{\text{M}}_{\alpha_2}$).

\subsection{Minimal Liouville gravity}
The MLG partition function is combined from the matter (M), Liouville (L) and ghosts (G) sectors
\be
Z_{\text{MLG}} = Z_{\text{M}} \cdot Z_{\text{L}} \cdot Z_{\text{G}}\;, 
\ee
Ghosts are just a free fermionic system with fields $B,C$ of conformal dimensions $(2,-1)$ respectively, with the corresponding antiholomorphic counterparts $\br{B}, \br{C}$ (its central charge is equal to $-26$).
All three sectors obey conformal symmetry and the zero total central charge condition,
$c_{\text{M}}+c_{\text{L}} = 26$, which follows from the Weyl invariance of the string action and ensures the nilpotency of the BRST-symmetry charge 
\begin{equation}\label{Q}
\mathcal{Q} \equiv \oint dz\, \left(C(T_L + T_M) + :C \pd C B:\right)(z)
\end{equation}
in MLG (and, similarly, for its antiholomorphic part $\br{\mathcal{Q}}$). In this paper we are interested in Yang-Lee series of minimal models in the matter sector, $r=2, r'=2p+1$, where $p$ is positive integer, see~\eqref{C_M}. In this case the central charge balance condition constraints the Liouville coupling constant to be $b=\sqrt{2/(2p+1)}$. 
  
We are interested in the physical operators, or BRST cohomologies $T_{m,n}$, constructed by dressing the minimal model primaries $\Phi_{m,n}(z)$\footnote{
For Yang-Lee series $m=1$ and Kac symmetry restricts $1\leq n\leq p$.}
with the Liouville (non-degenerate) primary exponential fields $V_a(z)$:
\be\label{T}
T_{m,n} = C\bar{C}\,U_{m,n}\;,\quad U_{m,n} \equiv \Phi_{m,n} V_{a_{m,-n}} \;,
\ee
where $a_{m,-n}$ is the solution of  the dimensional constraint 
\be\label{d-cond}
\Delta^{\text{M}}_{m,n}+\Delta^{\text{L}}_a=1\;.
\ee

Note that these cohomology classes can be represented either by inserting local operators $T_{m,n}$ at some point $z$, or by integrating the local density $U_{m,n}$ over the surface. Since the $\mathcal{Q}, \br{\mathcal{Q}}$-variation of $U_{m,n}$ is a total derivative, such an integral is BRST-invariant up to boundary terms. However, the second option becomes relevant only for higher multipoint correlators, which are not considered in this paper.\footnote{For constructing an invariant correlation number, it may be necessary to additionally dress $T_{m,n}$ with $B$-ghosts to provide a covariant correlator transformation, see the beginning of Section~\ref{sec:3}.}

\subsection{Ground ring in MLG and higher equations of motion}

 Apart from the tachyon physical operators~\eqref{T}, there exist a class of  ground ring operators \cite{wit1992},
constructed from degenerate Liouville primary fields,
\be\label{O}
O_{m,n} = H_{m,n} \br{H}_{m,n} \underbrace{\Phi_{m,n}V_{a_{m,n}}}_{\equiv \Theta_{m,n}} \;.
\ee
Here $H_{m,n}$ are polynomials of degree $mn-1$ of Virasoro generators and ghosts $B$ and $C$. Unlike the tachyons, the ground ring operators are built of the Virasoro descendants. The reason for their BRST invariance is that the result of the $\mathcal{Q}$ action is proportional to the singular descendants in the Liouville theory $D^{(L)}_{m,n} V_{a_ {m,n}}$ and in the minimal model $D^{(M)}_{m,n} \Phi_{m,n}$ (which we put to zero):
\begin{equation}
\mathcal{Q} O_{m,n} = \br{H}_{m,n} C  \mathcal{D}_{m,n}  \Theta_{m,n}\;,\quad \mathcal{D}_{m,n} = D^{(M)}_{mn} - (-1)^{mn} D^{(L)}_{mn}\;.
\end{equation} 
 No general form of $H_{m,n}$ is known, but it is easily found case by case when demanding the requirement above. We cite expressions for $H_{m,n}$ for the particular cases we will need in the next section (see e.g. \cite{imbimbo1992})
\begin{align}
&H_{1,2} = M_{-1} - L_{-1} + b^2 CB \label{h12}\;,    \\
&H_{1,4} = \text{(ghostless part)} + 9 b^4 BC (L_{-2}- M_{-2}) + \pd B\, C \left((12b^4- \frac{15b^2}{2}) L_{-1} - (12b^4+\frac{15b^2}{2}) M_{-1} \right)  \nonumber \\
&\qquad\quad + \frac{\pd^2 \,B C}{2} 9b^2(4b^4-1) \;, \label{h14} 
\end{align}
where by $M$ we denote Virasoro modes of the minimal model and by $L$ those of the Liouville theory. The ground ring operators have two important properties: independence of the correlators on their positions in a BRST-invariant environment and simple OPE with tachyons. These properties turn out to be useful in calculating MLG correlation numbers on the sphere \cite{Belavin:2005jy}. 

There is an important connection between the MLG gound ring operators and Zamolodchikov's higher equations of motion (HEM) \cite{highereoms2004}, which are a set of operator relations in Liouville QFT, that we briefly describe below. HEM refers to ``logarithmic'' operators of the form $\phi e^{2a \phi}$; such operators can be expressed as
\begin{equation}
V'_a = \frac{1}{2} \frac{\pd}{\pd a} V_a\;.
\end{equation}
We denote by $V'_{m,n}$ such a logarithmic operator evaluated at the value of the parameter corresponding to the degenerate dimension $a=a_{m,n}$. The simplest example is a Liouville field $\phi$ itself $V'_{1,1}$, which is subject to the ordinary Liouville equation of motion
\begin{equation}
\pd \br{\pd} \phi \equiv \pd \br{\pd} V'_{1,1} = \pi b \mu e^{2b\phi} \equiv \pi b \mu V_{1,-1}\;.
\end{equation}
In general, HEM is the relation between the field, which is obtained by applying the singular vector creating operator to the corresponding logarithmic field, and the primary field $V_{m,-n}$
\begin{equation}
  D^{(L)}_{m,n}  \br{D}^{(L)}_{m,n}  V'_{m,n} = B_{m,n} V_{m,-n}\;,
\end{equation}
where the constant
\begin{equation}
    B_{m,n} = (\pi \mu \gamma(b^2) b^{2-2b^2})^n \frac{\Upsilon'_b(2 a_{m,n})}{\Upsilon_b(2 a_{m,-n})}\;. \label{bmn}
\end{equation}
In the minimal Liouville gravity, these equations, together with some additional observations, lead to relations that are sometimes called the cohomological version of the HEM~\cite{Belavin:2005jy}. Since $V_{m,-n}$ in the RHS of HEM are precisely the dressing operators for the minimal model primaries $\Phi_{m,n}$ that appear in the MLG tachyons and these primaries are annihilated by operators $D^{(M)}_{m,n}$ and $\br{D}^{(M)}_{m,n}$, one gets 
\begin{equation}
U_{m,n} = V_{m,-n} \Phi_{m,n} = B_{m,n}^{-1} D^{(L)}_{m,n}  \br{D}^{(L)}_{m,n}  \underbrace{V'_{m,n} \Phi_{m,n}}_{\equiv \Theta'_{m,n}} = B_{m,n}^{-1} \mathcal{D}_{m,n} \br{\mathcal{D}}_{m,n} \Theta'_{m,n}\;.
\end{equation}
The RHS can be further rewritten using that
\begin{equation}
\mathcal{D}_{m,n} \br{\mathcal{D}}_{m,n} \Theta'_{m,n} = (\pd H_{m,n} - \mathcal{Q} R_{m,n}) (\br{\pd}\,\br{H}_{m,n} - \br{\mathcal{Q}}\, \br{R}_{m,n}) \Theta'_{m,n}\;,\quad R_{m,n} \equiv B_{-1} H_{m,n}\;. \label{chemfinal}
\end{equation}
%This can be understood by direct computation case by case. 
Thus, up to $\mathcal{Q}$-exact terms $U_{m,n}$ can be represented as a derivative of the operator $O'_{m,n} \equiv H_{m,n} \br{H}_{m,n} \Theta'_{m,n}$ which is a logarithmic counterpart of the ground ring operator $O_{m,n}$, eq.~\eqref{O}. 

Another form of higher equations of motion, which was first found in the context of super-MLG~\cite{Belavin:2008vc}, but is also valid in the bosonic case, reads
\begin{equation}
T_{m,n} \equiv C \br{C} V_{m,-n} \Phi_{m,n} = B_{m,n}^{-1}\mathcal{Q} \br{\mathcal{Q}} \left(O'_{m,n} \right)\;. \label{chemanotherform}
\end{equation}
It is easy to derive (\ref{chemfinal}) from this equation by applying $B_{-1}$ and commuting it with the BRST-charge.

\subsection{Some properties of CFT correlators on a torus}
In this section we collect some additional facts about CFTs on the torus (i.e., about  ghosts, MM and Liouville CFT) relevant to our problem.  
In what follows 
$\tau$ is the modular parameter of the torus 
and $q \equiv \exp (2\pi i \tau)$. 

First, we list the 
properties that are valid for all of the mentioned above  CFTs; most important are conformal Ward identities, which in particular express the correlators of the primary field $\Phi_\Delta$ with additional insertions of the stress-energy tensor $T$ in terms of the correlator without these insertions. For single and double stress-tensor insertions 
they read 
\begin{equation}
\langle  T(z) \Phi_\Delta(x) \rangle = \left[\Delta \left(\mathcal{P}(z-x) + 2\eta_1 \right) + (\zeta(z-x) + 2\eta_1 x) \pd_x + 2\pi i \frac{\pd}{\pd \tau} \right] \langle \Phi_\Delta (x) \rangle\;, \label{singletward}
\end{equation}
\begin{align}
&\langle  T(z) T(w) \Phi_\Delta(x) \rangle = \frac{c}{12} \mathcal{P}''(z-w) \langle \Phi_\Delta \rangle + \nonumber\\
&\quad +\left[2 (P(z-w) + 2 \eta_1) + (\zeta(z-w) + 2\eta_1 w)\pd_w \right] \langle T(w) \Phi_\Delta(x) \rangle + \nonumber \\
&\quad +  \left[\Delta (P(z-x) + 2 \eta_1) + (\zeta(z-x) + 2\eta_1 x) \pd_x \right] \langle T(w) \Phi_\Delta(x) \rangle + 2\pi i \frac{\pd}{\pd \tau} \langle T(w) \Phi_\Delta(x) \rangle\;. \label{doubletward}
\end{align}
Here elliptic $\zeta$-function and Weierstrass $\mathcal{P}$-function  behave as $\zeta(z)\sim 1/z$ and $\mathcal{P}(z)\sim1/z^2$ at $z=0$ 
(in this sense they are doubly periodic 
analogues of the corresponding terms arising in the conformal 
Ward identities on the sphere) and $\eta_1$ is a function of $\tau$
which is given by the following series expansion
\begin{equation}
\eta_1 = (2\pi)^2 \left[\frac{1}{24} + \sum \limits_{n=1}^\infty \frac{n q^n}{1-q^n} \right]\;.
\end{equation}
The derivation of the identities~\eqref{singletward} and~\eqref{doubletward}, as well as the precise definitions of $\zeta(z)$ and $\mathcal{P}(z)$, which we will not need in this paper, can be found in \cite{Eguchi:1986sb}.\footnote{Note that our normalization of $T$ differs from the one of \cite{Eguchi:1986sb} by a factor of $2\pi$.} 

Another common feature of the considered CFTs is that one-point functions of primary fields $f_\Delta(\tau) \equiv \langle \Phi_\Delta \rangle_\tau$ 
are modular forms of weight $\Delta$, which means that under transformation 
$\tau \to \tau+1$ they are invariant and under $\tau \to - \frac{1}{\tau}$ they transform 
as
\begin{equation}
f_\Delta (-\frac{1}{\tau}) = (\tau \br{\tau})^{\Delta} f_\Delta (\tau) \;.\label{modprop}
\end{equation}
This condition is important for a conformal theory to be consistent on higher genus Riemann surfaces~\cite{Sonoda:1988fq} (for the proof in case of minimal models see~\cite{Felder:1989vx}, where it is also shown that similar transformation rules take place for Virasoro 
descendant fields).

Let us now turn to the expressions for specific correlators in the QFTs we are interested in.
In the Hamiltonian formulation
torus correlators are represented as traces over the full set of states representing the spectrum of the theory.

We start with the ghost CFT. As pointed out for instance in \cite{polchinski_1998}, $BC$-system on the torus is to be quantized with the same boundary conditions as fields in the other sectors (i.e. periodic rather than antiperiodic in both directions), because it comes from the Faddeev-Popov determinant.
This means that the fields decompose in integer Fourier modes: $B(z) = \sum e^{2\pi i n z} B_{-n},\,C(z) = \sum e^{2\pi i n z} C_{-n}$ and the ghost correlator is to be understood as a trace with the insertion of the fermion parity operator $(-1)^F$. Operator coefficients $B_l,C_k$ satisfy $\lbrace B_l, C_k \rbrace = \delta_{l+k,0}$.

There are two degenerate vacua $\ket{\uparrow}, \ket{\downarrow}$ that form the two-dimensional representation of the subalgebra of zero modes $\lbrace B_0, C_0 \rbrace = 1$ (e.g. we have $C_0 \ket{\uparrow} = \ket{\downarrow}$, $B_0 \ket{\uparrow} = 0$) and are annihilated by $B_k, C_k$ with positive $k$. The basis states in the theory are built by applying to these two vacua modes $B_l, C_l$ with $l<0$; they can be labeled by two strictly  increasing Young diagrams $\lambda_1, \lambda_2$. 

For every basis state $B_{-\lambda_1} C_{-\lambda_2} \ket{\uparrow}$ there is a pair $B_{-\lambda_1} C_{-\lambda_2} \ket{\downarrow}$ built from another vacuum with the same $L_0$ eigenvalue, but different fermion number. In particular, this implies that $\langle \mathbb{I} \rangle_{gh}$ is zero. The same is valid for the antiholomorphic ghosts $\br{B}, \br{C}$. The simplest nonzero correlator is $\langle B(z)C(w) \br{B}(\br{z}) \br{C} (\br{w}) \rangle$; it is given by
\begin{equation}
    \langle B(z)C(w) \br{B}(\br{z}) \br{C} (\br{w}) \rangle = |\eta(q)|^4,\quad\eta(q) \equiv q^{1/24} \prod \limits_{n=1}^\infty (1-q^n) \;.\label{BCcorrelator}
\end{equation}
In what follows, we will often use its independence from the positions of the ghosts. In the Lagrangian formulation this is equivalent to the fact that the functional integral over the ghosts vanishes unless there are insertions to saturate the zero-modes, and there is one-zero mode for both $B$ and $C$ (they are just constant functions of $z$).

Minimal model one-point correlators (we will not need the higher multipoint ones) are just given by traces over Hilbert space, as mentioned before. Schematically,
\begin{equation}
  \langle \Phi \rangle_\tau = \sum \limits_{\Delta, \lambda, \br{\lambda}} q^{\Delta(\lambda) - \frac{c_L}{24}} \br{q}^{\Delta(\br{\lambda}) - \frac{c_L}{24}} \bra{\Delta, \lambda, \br{\lambda}} \Phi(z) \ket{\Delta, \lambda, \br{\lambda}}\;,
\end{equation}
where $\Delta$ are conformal dimensions and $\lambda$, $\br{\lambda}$ --- Young diagrams, labeling Virasoro descendants. The ratio $\bra{\Delta, \lambda, \br{\lambda}} \Phi(z) \ket{\Delta, \lambda, \br{\lambda}}/\bra{\Delta} \Phi(z) \ket{\Delta}$ is completely fixed by conformal symmetry and the series over $\lambda, \br{\lambda}$ for any $\Delta$ in the spectrum sums up to the toric conformal blocks (see e.g. \cite{Fateev:2009aw}).  Explicitly one gets
\begin{equation}
\langle \Phi_{1,k} \rangle = \sum \limits_{m=1}^p C_{m, (1,k)}^{(M) m} |q|^{2\Delta_{1,m}^\text{M} - \frac{1}{12} + \frac{(b^{-1}-b)^2}{2}} |F_M(\Delta^\text{M}_{1,k}, \Delta^\text{M}_{1,m}, q)|^2 \;.\label{1pfexmm}
\end{equation}
A similar formula is valid in the Liouville theory:
\begin{equation}
\langle V_a \rangle_\tau = \int \limits_\gamma \frac{dP}{4\pi} C^{(L)Q/2+iP}_{a, Q/2+iP} (q \br{q})^{-1/24 +P^2} \times |F_L(\Delta^\text{L}_a, \Delta^\text{L}_{Q/2+iP}, q)|^2 \;,\label{1pfexlio}
\end{equation}
where the ``diagonal'' structure constant is (it follows from (\ref{liouv3pf}))
\begin{equation}
C^{(L)Q/2+iP}_{a, Q/2+iP} = (\pi \mu \gamma(b^2) b^{2-2b^2})^{-a/b} \frac{\Upsilon(b) \Upsilon(2a) \Upsilon(2iP) \Upsilon(-2iP)}{\Upsilon^2(a) \Upsilon(a+2iP) \Upsilon(a-2iP)} \;.\label{diagscliouv}
\end{equation}
It has four series of poles in $P$ at $\pm 2iP = -m b - n b^{-1}-a $ and $\pm 2iP = Q + m b + n b^{-1}-a$. For $0<a<Q$  (as for instance for the Liouville fields with $a=a_{1,-n}$ which dress the minimal model primary fields) the integration contour $\gamma$ is just the real line; it ``separates'' the four series of poles in this case. For other $a$ (for example, degenerate ones), from the requirement of analyticity, as in the discussion of OPE, it is necessary to take into account the residues of the poles that cross the contour under analytic continuation.

\section{Calculation of torus one-point numbers}
\label{sec:3}
Because of the presence of ghost zero modes, in order to get a nonzero answer for a tachyon one-point correlator on the torus one needs to insert additional fields $B$ and $\br{B}$ together with $T_{m,n}$. Due to the properties described in the previous section, the correlator $\langle T_{m,n} B \br{B} \rangle_\tau$ is a modular form of weight $(2,2)$. It follows that we get a well-defined correlation number if we integrate this correlator over one-punctured torus moduli space, which is a fundamental domain of the $PSL(2,\mathbb{Z})$ action on upper half-plane.\footnote{Because of this property of the correlator, the correlation  number is independent of the choice of the fundamental domain.} 
Hence our main object is
\begin{equation}\label{one-point}
\int \limits_F d^2\tau\,\langle B \br{B} T_{1,n} \rangle_\tau = \int \limits_F d^2\tau\,\langle B \br{B} C \br{C} V_{1,-n} \Phi_{1,n} \rangle_\tau\;.
\end{equation}
In this section we illustrate the principal steps of the calculation on the simplest in the technical sense case $(m,n) = (1,2)$ and then describe how it generalizes to a more difficult case $(m,n) = (1,4)$, where we need to address additional questions.
\subsection{Case of $T_{1,2}$}
\subsubsection{Reduction to boundary terms}

First, note that the correlator is independent of ghost positions due to (\ref{BCcorrelator}). Correlators with the number of ghosts $B$ or $C$ less than one are zero, as well as correlators of the form $\langle \pd^k B\, \pd^l C \rangle$, $k\text{ or }l>0 $. Let's move $C$ to the same point where $V\Phi$ stands, and move $B$ to some other point $z$. Then we can rewrite the tachyon operator dressed by $C$ ghosts using HEM in the form (\ref{chemanotherform}). The explicit expression for the operator $H$ in the considered case is~(\ref{h12}).

One can commute the BRST operator with the remaining $B$-ghosts using $\lbrace Q, B(z) \rbrace = T(z)$, where $T = T_L + T_M + T_{gh}$ is the stress-energy tensor of the full theory, and discard $Q$-exact terms. This gives
\begin{equation}\label{uptoQexact}
\int d^2\tau\,\langle B(z) \br{B}(\br{z})\, \mathcal{Q} \br{\mathcal{Q}} \left(O'_{m,n} \right) \rangle_\tau = 
\int d^2\tau\,\langle T(z) \br{T}(\br{z}) O'_{m,n}  \rangle_\tau\;.
\end{equation}
The fact that $Q$-exact terms can be ignored can be understood as follows. In the Hamiltonian formulation, the torus average of BRST-exact operator $\mathcal{Q}(\dots)$ becomes the (super)trace over the Hilbert space of the form $\text{Tr}\,\left[(-1)^F q^{L_0} \br{q}^{\br{L}_0} \lbrace \mathcal{Q}, \dots \rbrace\right]$. Using the cyclic property of the trace and the fact that $\mathcal{Q}$ anticommutes with $(-1)^F$ and commutes with Virasoro modes, it formally evaluates to zero. However, using this argument in our case can be problematic since the action of the logarithmic operators is defined only on an extension of the MLG Hilbert space. As is known from the consideration of correlators in genus zero, $\mathcal{Q}$-descendants of the logarithmic operators do not necessarily give zero in a BRST-invariant environment. Noticing this subtle point, we nevertheless discard $\mathcal{Q}$-exact terms in~\eqref{uptoQexact} in order to move on.

Taking the derivative $\pd/ \pd a$ out of the integral over $\tau$, we are left with
\begin{equation}
\frac{1}{2} \frac{\pd}{\pd a}\int d^2\tau\,\langle T(z) \br{T}(\br{z})\, (L_{-1} - M_{-1} + b^2 BC)\br{ (\dots)} V_a \Phi_{1,2} \rangle = \frac{1}{2} \frac{\pd}{\pd a}\int d^2\tau\,\langle T(z) \br{T}(\br{z})\, b^4 B C \br{B} \br{C} V_a \Phi_{1,2} \rangle\;.
\end{equation}
The omitted terms are zero due to the absence of ghost operators either in the holomorphic or antiholomorphic sector (there are ghosts in $T_{gh}$, but at least one ghost in each term has a derivative with respect to $z$). Now we can use Ward identities (\ref{singletward}) to get rid of $T, \br{T}$ insertions, since other fields in the correlator are primary. This is one of the simplifications that occurs in the case $(1,2)$. From the Ward identities, we will also discard terms with derivatives over $w$ and $\br{w}$ on the basis of translational invariance of the one-point function.\footnote{For more discussion on this point see Section~\ref{oddk}.}

At $a = a_{1,2}$ the total dimension of the field $BC V_a \Phi_{1,2}$ is $0$. When differentiating over $a$, we either do not differentiate the prefactors coming from the Ward identities, which gives
\begin{equation}
(2\pi)^2 \frac{\pd^2}{\pd \tau \pd \br{\tau}} \langle BC \br{BC} V'_{1,2} \Phi_{1,2} \rangle_\tau \;,\label{cont10}
\end{equation}
or we differentiate the conformal dimension $\Delta_\Phi$ in one of these prefactors, then we get 2 terms
\begin{equation}
2\pi \Delta^{\text{L} '}_a \left(i \frac{\pd}{\pd \tau} \left[\br{\mathcal{P}} + 2 \br{\eta}_1 \right] - i \frac{\pd}{\pd \br{\tau}}  \left[\mathcal{P} + 2 \eta_1 \right]  \right) \langle BC \br{BC} V_{1,2} \Phi_{1,2} \rangle_\tau \;.\label{otherterm}
\end{equation}
Here $\pd /\pd \tau, \pd/ \pd \br{\tau}$ act on everything that stands on the right. We note that (\ref{otherterm}) has some expicit $z$-dependence, although this dependence was absent in the initial expression.

The listed above terms are given by derivatives on the moduli space, so they  can be reduced to the boundary contributions. The standard representation of the one-punctured torus moduli space is a region in $\tau$-plane $F_0 = \lbrace \text{Im }\tau>0,\,|\text{Re }\tau|<\frac{1}{2},\,|\tau|>1 \rbrace$ (see Figure \ref{fig:a}). The boundary components at $\tau_1 \equiv \text{Re }\tau = \pm 1/2$ are identified, as well as the arcs of the circle $|\tau|=1$, which are symmetric with respect to $\text{Re }\tau = 0$.\footnote{Note that gluing two arcs on the boundary component $|\tau| = 1$  can be done only if the integrand is modular invariant, which is not the case for $O'_{1,2}$. This causes an additional contribution, as described below.} Therefore, it is natural to think of the moduli space boundary as a horizontal segment  of length 1 at $\tau_2\equiv \text{Im }\tau = +\infty$. In this limit, most of the terms appearing in Liouville/MM/ghost correlators die out, being proportional to positive powers of $q = \exp (2\pi i \tau)$. Thus, we expect that in order to obtain an exact answer, we need to calculate only a finite number of terms in the correlator of order $q^0$. This is the main motivation for the upcoming calculation.

Let's address first the $z$ dependency problem in~\eqref{otherterm}. The solution is somewhat surprising: (\ref{otherterm}) is actually zero, because the average $\langle BC \br{BC} V_{1,2} \Phi_{1,2} \rangle$ is independent of $\tau, \br{\tau}$. There are two ways to illustrate this. The first one is algebraic, and consists of the following chain of relations 
(and analogous one for $\br{\tau}$)
\begin{align*}
&2\pi i \frac{\pd}{\pd \tau} \langle BC \br{BC} V_{1,2} \Phi_{1,2} \rangle  = \text{(Ward identities)} =  \langle T(z) BC \br{BC} V_{1,2} \Phi_{1,2} \rangle =  \text{(as discussed before)} \\
&= \langle T(z) O_{1,2}^{full} \rangle = \langle (\mathcal{Q} B(z)) O_{1,2}^{full} \rangle = \langle \mathcal{Q}( B(z) O_{1,2}^{full}) \rangle - \langle B(z) \cancel{(\mathcal{Q} O_{1,2}^{full})} \rangle = 0\;.
\end{align*}
This is an inversion of the argument presented earlier, analogous to the fact that correlators on the sphere do not depend on the positions of $\mathcal{Q}$-closed operators, if all other operators are BRST-closed. Another way is to calculate $\langle O_{1,2} \rangle$ brute force using explicit expressions for single point functions (\ref{1pfexmm}), (\ref{1pfexlio}) and toric conformal blocks \cite{Fateev:2009aw}; since we have a degenerate field in the Liouville sector, only discrete terms contribute to the Liouville one-point function (continuous contribution vanishes) and the expression for $\langle O_{1,2} \rangle_\tau$ is simple. We checked the independence of $q$ of the correlation function $\langle BC \br{BC} V_{1,2} \Phi_{1,2} \rangle_\tau$ up to the second order. In particular, this implies an interesting identity for toric conformal blocks with degenerate dimensions
\begin{equation}
\frac{\eta(q)^2}{q^{1/12}} \times F_L(\Delta^\text{L}_{1,2}, \Delta^\text{L}_{Q/2+ib/4}, q)  \times F_M(\Delta^\text{M}_{1,2}, \Delta^\text{M}_{1,p}, q) = 1\;.
\end{equation}
The brute force method is more difficult to adapt to the case of general $O_{1,k}$, but it can also follow from such non-trivial identities.

Now let's address the issue of modular non-covariance. As discussed in the previous section, one-point functions of primaries are modular forms (\ref{modprop}). For $\Delta = 0$, as, for example, for the ground ring operator $O_{1,2}$, this means that the correlator is modular invariant. For the logarithmic fields like $O_{1,2}'$, however, they transform non-covariantly. Taking $\Phi_\Delta = BC \br{BC} V_a \Phi_{1,2}$ and  differentiating both sides of (\ref{modprop}) with respect to $a$ and then setting $a$ such that $\Delta(a) = 0$, we get:
\begin{equation}
\frac{1}{2} \frac{\pd}{\pd a} f_\Delta (-\frac{1}{\tau}) = \frac{1}{2} \frac{\pd}{\pd a}  f_\Delta (\tau) + \frac{1}{2} \frac{\pd}{\pd a} \Delta^\text{L}_a \cdot \log (\tau \br{\tau}) f_\Delta(\tau)\;.
\end{equation}
We will denote the average $\langle O' \rangle_\tau \equiv \frac{1}{2} \frac{\pd}{\pd a} f_\Delta (\tau)$ as $F(r, \varphi)$, where $\tau = i r e^{i \varphi}$. Then the equation above can be rewritten as
\begin{equation}
F(1/r, -\varphi) = F(r,\varphi) + \frac{1}{2} \frac{\pd}{\pd a} \Delta^\text{L}_a \cdot \log (r^2) f_\Delta(\tau) \;.\label{needeq}
\end{equation}
 With this knowledge, let us rewrite~\eqref{cont10} as the boundary integral using Gauss theorem:
\begin{equation}
\int \limits_F dS\, \left(\vec{\nabla}, \vec{A} \right) = \int \limits_{\pd F} dl\, (\vec{n}, \vec{A})\;.
\end{equation}
The contribution from the boundary at infinity is
\begin{equation}
(2\pi)^2 \int \limits_F dS\, \underbrace{\frac{1}{4} \left(\vec{\nabla}, \vec{\nabla} \right)}_{\frac{\pd^2}{\pd \tau \pd \br{\tau}}} \langle O' \rangle = \frac{1}{4} (2\pi)^2 \int \limits_{-1/2}^{1/2} d\tau_1\,\frac{\pd}{\pd \tau_2}  \langle O' \rangle_{\tau_2 \to \infty} + \dots \label{cont1}
\end{equation}
and the integral over the arc segment of the boundary can be rewritten as 
\begin{equation}
- \frac{1}{4} (2\pi)^2 \int \limits_{-\pi/6}^{\pi/6} d\varphi\, \frac{\pd }{\pd r} F(r,\varphi)\mid_{r=1} = -\frac{1}{4} (2\pi)^2  \int \limits_{0}^{\pi/6} d\varphi\, \left[\frac{\pd }{\pd r} F(r,\varphi)+\frac{\pd }{\pd r} F(r,-\varphi) \right]\mid_{r=1} \;.
\end{equation}
There is no contribution from the vertical parts of the boundary with $\text{Re }\tau = \pm 1/2$ because the integrand is  invariant  under $\tau \to \tau + 1$. Differentiating (\ref{needeq}) with respect to $r$ and then setting $r=1$, we get
\begin{equation}
-\frac{\pd}{\pd r} F(r,-\varphi) \mid_{r=1} = \frac{\pd}{\pd r} F(r,\varphi) \mid_{r=1} + \frac{1}{2} \frac{\pd}{\pd a}\Delta^\text{L}_a \cdot 2 \cdot f_\Delta(\tau) 
%+ 0
\;.
\end{equation}
Therefore, the integral over the arc is defined by the expression
\begin{equation}
+ \frac{1}{4} (2\pi)^2  \frac{1}{2} \frac{\pd}{\pd a}\Delta^\text{L}_a \cdot 2 \int \limits_0^{\pi/6} d\varphi\, \langle O \rangle_{\tau = i \exp (i\varphi)}  \;.\label{cont3}
\end{equation}
Using the fact that $\langle O_{1,2} \rangle_\tau$ does not depend on $\tau$, which we stated earlier, eq.~\eqref{cont10} can be finally rewritten as
\begin{equation}
\frac{1}{4} (2\pi)^2 \int \limits_{-1/2}^{1/2} d\tau_1\,\frac{\pd}{\pd \tau_2}  \langle O' \rangle_{\tau_2 \to \infty} + \frac{1}{24} (2\pi)^3 \cdot  \frac{1}{2} \frac{\pd}{\pd a}\Delta^\text{L}_a \cdot  \langle O \rangle \;.\label{rewrcont3}
\end{equation}

\subsubsection{Calculation of $\langle O_{1,2} \rangle_\tau$ and $\langle O'_{1,2} \rangle_\tau,\,\tau_2 \to \infty$}
 In this section we will discuss how to calculate $\langle O \rangle$ and $\langle O' \rangle$. First, the average in the  minimal model is given as a sum over $p$ conformal families (\ref{1pfexmm})
\begin{equation}
\langle \Phi_{1,2} \rangle = \sum \limits_{m=1}^p C_{m, (1,2)}^{(M) m} |q|^{2\Delta^\text{M}_{1,m} - \frac{1}{12} + \frac{(b^{-1}-b)^2}{2}} |F_M(\Delta^\text{M}_{1,2}, \Delta^\text{M}_{1,m}, q)|^2 \;.\label{minmodsum}
\end{equation}
In the case of $\Phi_{1,2}$, the diagonal structure constant $C_{m, (1,2)}^{(M) m} $ is only nonzero for $m=p$, because $\Phi_{1,p}$ and $\Phi_{1,p+1}$ are identified. In this particular case it is equal to
\begin{equation}
 C_{p, (1,2)}^{(M) p+1}= \left( \frac{\gamma(2-2b^2) \gamma(1- p b^2)}{\gamma(1-b^2) \gamma(2 - b^2(p+1))} \right)^{1/2} \;.
\end{equation}
The correlator of the ghosts is given by (\ref{BCcorrelator}), it can be written as a series expansion in $q$, which starts with $|q|^{1/6}$. The Liouville part is given by (\ref{1pfexlio}).

The normalization of the conformal blocks $F_L$ and $F_M$ is such that they start with $1$ when expanded into series in $q$. Let us look at leading terms in the series expansion in $q$ of the total correlator. Substituting $\Delta^\text{M}_{1,m} = \Delta^\text{M}_{1,p}$, we get that it is of order $|q|^{(b^2/8+2 P^2)}$. If $P$ is real, even this leading term tends to zero on the boundary of moduli space. Therefore, the continuous contribution to the Liouville correlator vanishes. The Liouville structure constant (\ref{diagscliouv}) also linearly tends to zero when $a \to a_{1,2} = - \frac{b}{2}$. There are, however, discrete contributions from the poles of structure constant (which have imaginary $P$) that cross the integration contour when $a \to a_{1,2}$. If two of $\Upsilon$-functions in the denominator of the structure constant become zero when $a \to a_{1,2}$, the residues do not vanish in the limit. There are two such poles $P = \pm \frac{ia}{2} \to \mp \frac{ib}{4}$. Note that for such Liouville momenta both contributions to correlator from these poles are of order $|q|^0$ in the limit. Moreover, they are equal, because they are connected via reflection $P \to -P$. 
We now have an expression for the correlator
\begin{equation}
\langle O_{1,2} \rangle_{\tau_2 \to \infty} = b^4 \times C_{p,(1,2)}^{(M) p+1} \times \frac{2\pi i \times 2}{4\pi} \text{Res}_{P = ia/2} C_{a,Q/2+iP}^{(L) Q/2+iP} \mid_{a=a_{1,2}}\;.
\end{equation}
Let's see how this consideration is modified for the logarithmic field $\langle O' \rangle$. We are interested in the contributions proportional to $\tau_2 = \text{Im }\tau$, because in the boundary integral we have to differentiate with respect to $\tau_2$. This can only appear when we keep the $a$-dependence in the prefactor $|q|^{2P^2(a)}$ in discrete terms and differentiate it with respect to $a$ (differentiating the structure constant would instead give zero when we take the limit, so we leave it untouched). In our case $P(a) = \pm ia/2$ and the differentiation gives
\begin{align}
&\frac{\pd}{\pd \tau_2} \langle O'_{1,2} \rangle_{\tau_2 \to \infty}  = \lim \limits_{a \to a_{1,2}} \left[ \frac{\pd}{\pd \tau_2}  \frac{1}{2} \frac{\pd}{\pd a} \exp \left(2\pi \tau_2 a^2/2 \right)  \langle O_{1,2} \rangle + O(a-a_{1,2}) \right] = \nonumber \\
& = \pi a_{1,2}  \langle O_{1,2} \rangle = - \frac{\pi b}{4} \cdot 2   \langle O_{1,2} \rangle  \;.\label{oprime12ando12}
\end{align}
Now, using
\begin{equation}
\frac{1}{2} \frac{\pd}{\pd a} \Delta^\text{L}_a = \frac{Q}{2} - a = (a = a_{1,k} = - \frac{(k-1) b}{2}) = \frac{b^{-1}}{2} + \frac{kb}{2} = \frac{b}{4} (2p +1 + 2k)\;,
\end{equation}
the two terms in (\ref{rewrcont3}) together  finally yield 
\begin{equation}
 \frac{(2\pi)^3}{24} \frac{b}{4}(2p+1+4) \langle O_{1,k} \rangle - \frac{(2\pi)^2}{4} \frac{4\pi b}{8} \langle O_{1,2} \rangle = \frac{(2\pi)^3 b}{96} (2p-1)  \langle O_{1,2} \rangle  \;.
\end{equation}

\subsection{Case of $T_{1,4}$}
We now want to consider the case when the descendants of the fields $V$ and $\Phi$ appear in the operator $H_{m,n}$ and become relevant, in particular, the case $(1,4)$, see eq.~\eqref{h14}.  We will use the Ward identities on the torus~\eqref{singletward} and~\eqref{doubletward}, discarding terms with derivatives over $z$, as before. To make sure everything goes according to the same scenario as for the $(1,2)$ case, we need to check the following statements
\begin{enumerate}
\item The average of the ground ring operator $\langle O_{1,4} \rangle$ is $\tau$ and $\br{\tau}$ independent. As before, taking into account that $O_{1,4}$ is BRST-closed, it is enough to prove that the derivative with respect to $\tau$ ($\br{\tau}$) can be obtained by inserting $\mathcal{Q}$-exact stress energy tensor: $2\pi i \frac{\pd}{\pd \tau} \langle O_{1,4} \rangle = \langle T(z)  O_{1,4} \rangle$.
\item Given the first point, we need to prove that for the logarithmic operator
\begin{equation}
\frac{\pd}{\pd a} \langle T(z) \br{T} (\br{z}) H_{1,4} \br{H}_{1,4} V_a \Phi_{1,4} \rangle \mid_{a=a_{1,4}} = (2\pi)^2 \frac{\pd^2}{\pd \tau \pd \br{\tau}} \frac{\pd}{\pd a} \langle H_{1,4} \br{H}_{1,4} V_a \Phi_{1,4}  \rangle \mid_{a=a_{1,4}}\;.
\end{equation}
\end{enumerate}
To this end we first consider the average of the form (focusing on the holomorphic sector)
\begin{align}
&\langle T(z) H_{1,4} V_a \Phi_{1,4} \rangle = 9b^4 \langle T(z) BC (L_{-2} - M_{-2}) V_{a} \Phi_{1,4} \rangle + \nonumber \\
&+ \langle T(z) \pd B\,C (\#L_{-1} - \# M_{-1}) V_a \Phi_{1,4} \rangle + \frac{9b^2(4b^4-1)}{2} \langle T(z) \pd^2 B\,C V_a \Phi_{1,4} \rangle\;. \label{oaaverage}
\end{align}
Here we dropped the ghostless part in $H_{1,4}$, because, as before, the corresponding expectation value vanishes, since the ghost zero modes are not saturated.
\subsubsection{Reduction to boundary terms}
\paragraph{Terms with ghost derivatives in $H$.} Let us first consider the second and third terms in~\eqref{oaaverage}. Since the two-point ghost function $\langle \pd^k B\,C \rangle = 0$, for $k={1,2}$, the contribution from $(T_L + T_M)(z)$ in these terms vanishes because of the ghost sector. However, the four-point function involving  the derivatives of the ghosts (and thus the ghost expectation values $\langle T_{gh}(z) \pd^k B \,C \rangle$ is non-zero. In fact, from the Ward identity, using the coordinate invariance of the two-point function without derivatives, we get
\begin{align}
&\pd_x^k \langle T_{gh}(z) B(x) C(0) \rangle = \pd_x^k \left[2 (\mathcal{P}(z-x) + 2 \eta_1) - (\mathcal{P}(z) + 2 \eta_1) + 2\pi i \frac{\pd}{\pd \tau} \right] \langle B(x) C(0) \rangle \nonumber = \\
&= 2(-1)^k \mathcal{P}^{(k)}(z-x) \langle BC \rangle\;.
\end{align}
Taking into account these facts, we conclude that the second term in (\ref{oaaverage}) vanishes because of translational invariance (now in MM and Liouville sector) and the third one contributes
\begin{equation}
9b^2(4b^4-1) \mathcal{P}''(z) \langle BC V_a \Phi_{1,4} \rangle\;.
\end{equation}

\paragraph{The first term. } For the term without ghost derivatives, we represent Virasoro modes as contour integrals of the stress-energy tensor: $L_{-2} = \frac{1}{2\pi i} \oint \frac{dw}{w} T_L(w)$ and rewrite it as
\begin{equation}
\frac{1}{2\pi i} \oint \frac{dw}{w} \langle (T_L + T_M + T_{gh})(z) BC (T_L(w) - T_M(w)) V_{a} \Phi_{1,4}(0)\rangle\;.
\end{equation}
Let us first list the terms that do not need Ward identities for double $T$ insertion:
\begin{align}
&T_{gh} (T_L-T_M):~~\left[1 (\mathcal{P}(z) + 2\eta_1) + 2 \pi i\frac{\pd}{\pd \tau_{gh}} \right] \langle BC (L_{-2} - M_{-2}) V_{a} \Phi_{1,4}(0)\rangle\;,\\
&T_{L} (-T_M):~~\left[\Delta^\text{L}_a (\mathcal{P}(z) + 2\eta_1) + 2\pi i \frac{\pd}{\pd \tau_L} \right] \langle BC (- M_{-2}) V_{a} \Phi_{1,4}(0)\rangle\;, \\
&T_{M} T_L:~~\left[\Delta^{\text{M}}_{1,4} (\mathcal{P}(z) + 2\eta_1) + 2 \pi i \frac{\pd}{\pd \tau_M} \right] \langle BC L_{-2} V_{a} \Phi_{1,4}(0)\rangle \;.
\end{align}
We denote by partial $\tau$-derivative with a subscript the derivative acting only on the corresponding factor in the (factorized) one-point correlator. The total derivative is a sum $\frac{\pd}{\pd \tau} = \frac{\pd}{\pd \tau_{gh}} + \frac{\pd}{\pd \tau_{L}} + \frac{\pd}{\pd \tau_{M}}$. 
Now we write one of the remaining 2 terms, the one with double $T_L$ insertion, using (\ref{doubletward}) 
\begin{align}
&\frac{c_L}{12} \mathcal{P}''(z) \langle BC V_a \Phi_{1,4} \rangle + \oint \frac{dw}{(2\pi i) w} \left[2 (\mathcal{P}(z-w) + 2 \eta_1)  + (\zeta(z-w) +  2 \eta_1 w) \pd_w \right] \times \nonumber \\
& \times \left[\Delta^\text{L}_a (\mathcal{P}(w) + 2 \eta_1) + 2 \pi i \frac{\pd}{\pd \tau_L} \right] \langle BC V_a \Phi_{1,4} \rangle  + \left[\Delta^\text{L}_a (\mathcal{P}(z) + 2 \eta_1) + 2 \pi i \frac{\pd}{\pd \tau_L}  \right]  \langle BC L_{-2} V_a \Phi_{1,4} \rangle\;. \label{doubletl}
\end{align}
Let us carefully extract the terms of order $1/\omega$ which contribute to the integral in the second term of the equation above. Taking into account that $\mathcal{P}(w) \sim \frac{1}{\omega^2},\,\omega \to 0$, we obtain
\begin{align}
& \left[2 (\mathcal{P}(z-w) + 2 \eta_1)  + (\zeta(z-w) +  2 \eta_1 w) \pd_w \right] \left[\Delta^\text{L}_a (\mathcal{P}(w) + 2 \eta_1) + 2 \pi i \frac{\pd}{\pd \tau_L} \right] = \nonumber \\
&= 2 \left(\mathcal{P}(z) + 2 \eta_1 - w \mathcal{P}'(z) + \frac{w^2}{2} \mathcal{P}''(z)\right)  \left(2 \eta_1 \Delta_a + 2 \pi i \frac{\pd}{\pd \tau_L} + \frac{\Delta^\text{L}_a}{w^2} + \dots \right) +  \nonumber \\
&+ \left(\zeta(z) +  (2 \eta_1 - \zeta'(z)) w + \frac{w^2}{2} \zeta''(z) - \frac{w^3}{6} \zeta'''(z) \right) \left(- \frac{2\Delta^\text{L}_a}{w^3} + \dots \right) =  \nonumber \\
&= 2 \left(\mathcal{P}(z) + 2 \eta_1 \right) \left(2 \eta_1 \Delta^\text{L}_a + 2 \pi i \frac{\pd}{\pd \tau_L} \right) + \frac{2\Delta^\text{L}_a}{3} \mathcal{P}''(z)\;.
\end{align}
We used that $\zeta'''(z) = - \mathcal{P}''(z)$. With single-$T$ insertion Ward identities
\begin{equation}
 \langle BC L_{-2} V_a \Phi_{1,4} \rangle = \left(2 \eta_1 \Delta^\text{L}_a + 2 \pi i \frac{\pd}{\pd \tau_L} \right) \langle BC V_a \Phi_{1,4} \rangle\;.
\end{equation}
the expression (\ref{doubletl}) can be rewritten as follows:
\begin{equation}
\left(\frac{c_L}{12} + \frac{2\Delta^\text{L}_a}{3} \right) \mathcal{P}''(z) \langle BC V_a \Phi_{1,4} \rangle + \left[ (\Delta^\text{L}_a + 2) (\mathcal{P}(z) + 2 \eta_1) + 2\pi i \frac{\pd}{\pd \tau_L} \right]  \langle BC L_{-2} V_a \Phi_{1,4} \rangle\;.
\end{equation}
Similarly, the term coming from $T_M(z) T_M(w)$ gives
\begin{equation}
\left(-\frac{c_M}{12} - \frac{2\Delta^{\text{M}}_{1,4}}{3} \right) \mathcal{P}''(z) \langle BC V_a \Phi_{1,4} \rangle + \left[ (\Delta^{\text{M}}_{1,4} + 2) (\mathcal{P}(z) + 2 \eta_1) + 2\pi i \frac{\pd}{\pd \tau_M} \right]  \langle BC (-M_{-2}) V_a \Phi_{1,4} \rangle\;.
\end{equation}
Summing all the contributions, we obtain
\begin{align}
&\left[9b^4 \left(\frac{c_L - c_M}{12} + \frac{2 (\Delta^\text{L}_a -\Delta^{\text{M}}_{1,4})}{3} \right) + 9b^2(4b^4-1) \right] \mathcal{P}''(z) \langle BC V_a \Phi_{1,4} \rangle + \nonumber \\
&+  (\Delta^\text{L}_a + \Delta^{\text{M}}_{1,4} + 3)  (\mathcal{P}(z) + 2 \eta_1)  \langle BC (L_{-2} -M_{-2}) V_a \Phi_{1,4} \rangle + \nonumber \\
&+  2\pi i \frac{\pd}{\pd \tau}  \langle BC (L_{-2} -M_{-2}) V_a \Phi_{1,4} \rangle\;.
\end{align}
As expected, the prefactors in the first and second lines vanish at $a=a_{1,4}$ (for the expectation value  $\langle T(z) O_{1,4} \rangle$), leaving only the derivative with respect to the modular parameter. This proves the $\tau$ and $\br{\tau}$-independence of $\langle O_{1,4} \rangle$. To prove it for the logarithmic operator, we use the following brief notation:
\begin{align}
&\left[9b^4 \left(\frac{c_L - c_M}{12} + \frac{2 (\Delta^\text{L}_a -\Delta^{\text{M}}_{1,4})}{3} \right) + 9b^2(4b^4-1) \right] \mathcal{P}''(z) = \beta(a-a_{1,4})\;, \nonumber \\
&(\Delta^\text{L}_a + \Delta^{\text{M}}_{1,4} + 3)  (\mathcal{P}(z) + 2 \eta_1)  = \gamma (a-a_{1,4})\;.
\end{align}
We also use analogous notation to describe the antiholomorphic dependence. Then we have
\begin{align}
&\langle T(z) \br{T}(\br{z}) H_{1,4} \br{H}_{1,4} V_a \Phi_{1,4} \rangle = \beta(a-a_{1,4}) \langle \br{T} BC \br{H}_{1,4} V_a \Phi_{1,4} \rangle + \nonumber \\
&+ \left(2\pi i \frac{\pd}{\pd \tau} + \gamma (a-a_{1,4})  \right)  \langle \br{T} H_{1,4} \br{H}_{1,4} V_a \Phi_{1,4} \rangle =  \nonumber\\
&= \beta(a-a_{1.4}) \left[\br{\beta}(a-a_{1,4}) \langle BC \br{BC} V_a \Phi_{1,4} \rangle +  \left(-2\pi i \frac{\pd}{\pd \br{\tau}} + \br{\gamma} (a-a_{1,4})  \right) \langle BC \br{H}_{1,4} V_a \Phi_{1,4} \rangle \right] + \nonumber \\
&+ \left(-2\pi i \frac{\pd}{\pd \br{\tau}} + \br{\gamma} (a-a_{1,4}) \right)  \left(2\pi i \frac{\pd}{\pd \tau} + \gamma (a-a_{1,4})  \right)  \langle H_{1,4} \br{H}_{1,4} V_a \Phi_{1,4} \rangle + \nonumber\\
&+  \br{\beta} (a-a_{1,4}) \left(2\pi i \frac{\pd}{\pd \tau}+  \gamma (a-a_{1,4})  \right)   \langle H_{1,4}\br{BC} V_a \Phi_{1,4} \rangle \;.
\end{align}
We see that after differentiating with respect to $a$ and substituting $a=a_{1,4}$ the only nonzero term\footnote{
All other terms either contain higher powers of $(a-a_{1,4})$, or are zero because  the expectation values of the ground ring elements do not depend on $\tau$.} is
\begin{equation}
\frac{1}{2} \frac{\pd}{\pd a} \langle T(z) \br{T}(\br{z}) H_{1,4} \br{H}_{1,4} V_a \Phi_{1,4} \rangle = (4\pi^2) \frac{\pd^2}{\pd \tau \pd \br{\tau}}  \frac{1}{2} \frac{\pd}{\pd a}  \langle H_{1,4} \br{H}_{1,4} V_a \Phi_{1,4} \rangle =  (4\pi^2) \frac{\pd^2}{\pd \tau \pd \br{\tau}} \langle O'_{1.4} \rangle\;.
\end{equation}

%%%
\subsubsection{Calculation of $\langle O_{1,4} \rangle_\tau$ and $\langle O'_{1,4} \rangle_\tau,\,\tau_2 \to \infty$}
Similarly to the $(1,2)$ case, here we have 2 contributions to the sum in the matter correlator (corresponding to $m=p \sim p+1$ and $m=p-1 \sim p+2$) or 4 Liouville discrete terms (two pairs connected by reflection) with Liouville momenta $P= \pm \frac{ib}{4}$ and $P = \pm \frac{3ib}{4}$. When one multiplies the OPE coming from both sectors, there appear 4 bilinear combinations of conformal blocks, such that the leading power of $q$ is not necessarily zero. Namely, we have
\begin{align*}
& m = p-1:\quad P = \pm \frac{ib}{4}- \langle \dots \rangle  \sim |q|^{b^2}\;,\quad P = \pm \frac{3ib}{4} - \langle \dots \rangle \sim |q|^{0} \\
& \,\,\, m= p:\quad P = \pm \frac{ib}{4} - \langle \dots \rangle \sim |q|^{0}\;,\quad P = \pm \frac{3ib}{4} - \langle \dots \rangle \sim |q|^{-b^2} 
\end{align*}
%One of these 4 terms is particularly worrying since it grows when $|q| \to 0$. However, 
The two contributions which are not of the order $q^0$ actually vanish, because there is an additional coefficient coming from $L,M$ Virasoro modes in (\ref{h14}). From the conformal Ward identity (\ref{singletward}) it follows that  
\begin{equation}
\langle (L_{-2} - M_{-2}) V_{1,4} \Phi_{1,4} \rangle = \left[2 \eta_1 \left(\Delta^\text{L}_{1,4} - \Delta^{\text{M}}_{1,4} \right) + 2 \pi i \left(\frac{\pd}{\pd \tau_L} - \frac{\pd}{\pd \tau_M} \right) \right] \langle V_{1,4} \Phi_{1,4} \rangle \;.\label{conseqtward}
\end{equation}
Differentiating the leading terms of the four considered contributions with respect to $\tau$  reduces to an additional numeric factor, which depends on $m$ and $P$. We thus replace the result of the action of the  derivative operators with this numeric factor as follows:
\begin{equation}
\frac{\pd}{\pd \tau_L} - \frac{\pd}{\pd \tau_M} \to 2\pi \left(\Delta^{\text{M}}_m - \Delta_P^L + \frac{c_L-c_M}{24} \right)\;.
\end{equation}
Moreover (again, up to leading order in $q$) we have $\eta_1 \approx \frac{1}{24} (2\pi)^2$, then in total RHS of (\ref{conseqtward}) can be rewritten in the following way
\begin{equation}
(2\pi)^2 \left[\frac{1}{12}\left(\Delta^\text{L}_{1,4} - \Delta^{\text{M}}_{1,4} \right) +  \left(\frac{(b^{-1}-b)^2}{4} + \Delta^{\text{M}}_m - P^2 \right) \right] (1+O(q))\langle V_{1,4} \Phi_{1,4} \rangle\;.
\end{equation}
It is easy to verify that for the cases $(m=p-1, P = \pm \frac{ib}{4})$ and $(m=p, P = \pm \frac{3ib}{4})$ the prefactors are zero, and for the other two combinations it simplifies to $\pm (2\pi)^2 \frac{b^2}{2}$. Since such a factor comes from both holomorphic and antiholomorphic Virasoro modes, the sign is irrelevant and in the terms of interest we can replace Virasoro modes $(L_{-2} - M_{-2})$ with a numeric factor $(2\pi)^4 \frac{b^4}{4}$. To calculate the expectation value $\langle O_{1,4} \rangle$ it remains to calculate the product of the degenerate structure constants corresponding to two remaining choices of $m$ and $p$.

 An interesting observation is that all 4 nonzero terms (if the Liouville contributions associated with reflection are taken into account separately) turn out to be equal. The contributions to the correlator from $m=p$ and $m=p-1$ with the corresponding Liouville discrete terms each give half of $\langle O_{1,4} \rangle$. This is because we have an identity 
\begin{equation}
C_{p,(1,4)}^{(M) p+1} \text{Res}_{P = ib/4} C_{a_{1,4},Q/2+iP}^{(L) Q/2+iP} = C_{p-1,(1,4)}^{(M) p+2} \text{Res}_{P = 3ib/4} C_{a_{1,4},Q/2+iP}^{(L) Q/2+iP}\;.
\end{equation}
When considering the logarithmic operator $O'_{1,4}$ instead of $O_{1,4}$, two pairs of reflection-related discrete terms after differentiation with respect to $a$ are additionally multiplied by corresponding Liouville momenta (up to a factor of $i/2$ they are equal to $a_{1,4} = \frac{-3b}{2}$ and $a_{1,4}+b = \frac{-b}{2}$), so we would have
\begin{equation}
\frac{\pd}{\pd \tau_2} \langle O'_{1,4} \rangle_{\tau_2 \to \infty}  = - \pi  \cdot \frac{1}{2}\left(\frac{3b}{2} \langle O_{1,4} \rangle + \frac{b}{2} \langle O_{1,4} \rangle  \right) = - \frac{\pi b}{4} \cdot 4 \langle O_{1,4} \rangle \;.\label{oprime14ando14}
\end{equation}
We proved above that the expression in the form (\ref{rewrcont3}) is also valid in case $(1,4)$. We thus have the final answer for tachyon one-point function
\begin{equation}
\langle T_{1,4} \rangle = \frac{(2\pi)^3}{24} \frac{b}{4}(2p+1+8) \langle O_{1,4} \rangle - \frac{(2\pi)^2}{4} \frac{8\pi b}{8}  \langle O_{1,4} \rangle = \frac{(2\pi)^3 b}{96} (2p-3)   \langle O_{1,4} \rangle\;.
\end{equation}

\section{Discussion}
\label{sec:concl}

\subsection{General case $T_{1,2k}$ and comparison with matrix model}
Building on the observations for the example of $O_{1,4}$ about the fact that all 4 terms in the sum for $\langle O \rangle$ turn out equal, as well as extrapolating formulas (\ref{oprime14ando14}) and  (\ref{oprime12ando12})   we conclude by conjecturing the following generalized formulas
%that in the natural normalization we would have
\begin{equation}
\langle O_{1,2k} \rangle = 2k\;,\quad\langle O'_{1,2k} \rangle = - \frac{\pi b}{4} \cdot 2k \langle O_{1,2k} \rangle = -\pi b k^2\;.
\end{equation}
Then, if all the steps leading to (\ref{rewrcont3}) are valid, we would get for even $k$
\begin{equation}
\langle T_{1,k} \rangle = \frac{(2\pi)^3}{24} \frac{b}{4}(2p+1+2k) \langle O_{1,k} \rangle - \frac{(2\pi)^2}{4} \frac{2\pi b}{8} k \langle O_{1,k} \rangle = \frac{(2\pi)^3 b}{96} (2p+1-k)  \langle O_{1,k} \rangle \sim k(2p+1-k) \;.
\label{T1kgeneral}
\end{equation}
This is precisely the matrix model result  obtained in~\cite{Belavin:2010pj}.

\subsection{Comments on the case $T_{1,k}$ with odd $k$}
\label{oddk}
In our examples, we ignored terms with derivatives over $w$ and $\br{w}$ on the basis of translational invariance of the one-point function. It is known from considering ordinary Liouville EOM that this can be a subtle issue for logarithmic operators --- we have a non-zero average of operator $\pd \br{\pd} \phi$ given by derivatives of the Liouville field. However, in that case, zero arising from the translational invariance cancels with the singularity of the diagonal structure constant $C_{0, Q/2+iP, Q/2-iP}$. There exists a regularisation in which this statement can be made precise; we describe it in Appendix \ref{eomliouv}.  We do not have such a singularity in the expectation value of $V_{1,2}$, as well as of any $V_{1,2k}$, so we do not expect such problems. However, for $\langle T_{1,k} \rangle$ with odd $k$ we have to treat carefully the terms containing derivatives with respect to the position of the tachyon.

For odd $k$ tachyons, there is also another problem (or another manifestation of the problem described above). The simplest demonstration can be done for $k=1$; in this case we have $H_{1,1} = 1$, containing no Virasoro modes nor ghosts. After using HEM and commuting the BRST-charge with the field $B(z)$, we obtain the following correlator:
\begin{equation}
\langle T(z) \br{T}(z) V'_{1,1} \Phi_{1,1} \rangle = \langle T(z) \br{T}(z) \phi\, \mathbb{I}_M \rangle\;.
\end{equation}
It is easy to see that the average in the ghost sector vanishes for this correlator. The expression we started with is nonzero (in fact, this particular tachyon one-point function was computed analytically long before with another methods in \cite{Bershadsky:1990xb}), which is an obvious contradiction. However, the Liouville part of this correlator $\langle \phi \rangle$ is in fact infinite (even more so than the torus partition function  $\langle \mathbb{I}_L \rangle$, divergent because of continuity of the spectrum of the theory). It is likely that the zero from the ghost sector may somehow cancel with this infinity, but we have not yet found a regularization that would allow to demonstrate this cancellation explicitly.

Despite the caveats above, the correlation numbers $T_{1,k}$ with odd $k$  can still be calculated with the method described in this paper, if before applying HEM we use that tachyons related by reflection in MLG are idenfitifed: $T_{1,k} = T_{1,2p+1-k}$. This reflection interchanges even and odd $k$, so after using this relation, the problems described above become irrelevant. Note that the expression obtained with our methods (\ref{T1kgeneral}) evidently respects the reflection symmetry.

\subsection{Other choices of fundamental domain}

In this subsection we want to check that our construction is consistent with the requirement of modular symmetry for the tachyon amplitudes~\eqref{one-point}. The partition function is modular invariant, which can be checked explicitly using the transformation properties of the ghost correlation function $\langle B\bar{B} C\bar{C}\rangle$ and modular properties of the characters. The correlation functions with insertions of the vertex operators are modular covariant and transform according to the transformation law for the vertex operators (see eq.\eqref{modprop}). This condition combined with the dimensional constraint~\eqref{d-cond} leads to the modular invariance of the amplitude~\eqref{one-point}.
The question is whether our regularization procedure used in Sections~\ref{sec:2} and~\ref{sec:3} is consistent with this requirement.

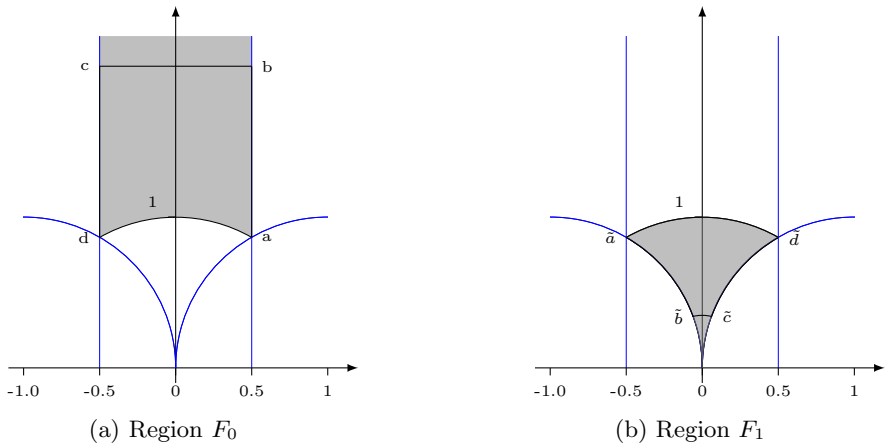
\begin{figure}[H]
\centering

\subfloat[Region $F_0$]{\label{fig:a}
\pgfmathsetmacro{\myxlow}{-1}
\pgfmathsetmacro{\myxhigh}{1}
\pgfmathsetmacro{\myiterations}{1}

\begin{tikzpicture}[scale=2]
           \draw[very thin, blue] (0.5,0)-- (0.5,2.2);
          \draw[very thin, blue] (-0.5,0)-- (-0.5,2.2);
 
    \draw[-latex](\myxlow-0.1,0) -- (\myxhigh+0.2,0);
    \pgfmathsetmacro{\succofmyxlow}{\myxlow+0.5}
    \foreach \x in {\myxlow,\succofmyxlow,...,\myxhigh}
    {   \draw (\x,0) -- (\x,-0.05) node[below,font=\tiny] {\x};
    }
    \foreach \y  in {1}
    {   \draw (0,\y) -- (-0.05,\y) node[above left,font=\tiny] {\pgfmathprintnumber{\y}};
   }
    \draw[-latex](0,-0.1) -- (0,2.4);
    \begin{scope}   
        \clip (\myxlow,0) rectangle (\myxhigh,1.1);
        \foreach \i in {1,...,\myiterations}
        {   \pgfmathsetmacro{\mysecondelement}{\myxlow+1/pow(2,floor(\i/3))}
            \pgfmathsetmacro{\myradius}{pow(1/3,\i-1}
            \foreach \x in {-2,\mysecondelement,...,2}
            {   \draw[very thin, blue] (\x,0) arc(0:180:\myradius);
                \draw[very thin, blue] (\x,0) arc(180:0:\myradius);
            }   
        }
    \end{scope}
    \begin{scope}
        \begin{pgfonlayer}{background}
            \clip (-0.5,0) rectangle (0.5,2.7);
            \clip   (1,2.2) -| (-1,0) arc (180:0:1) -- cycle;
            \fill[gray,opacity=0.5] (-1,-1) rectangle (1,3);
        \end{pgfonlayer}
    \end{scope}
    
     \draw[name path=path1] (-0.5,.865) arc(120:60:1) node[at start,left,font=\tiny] {d};
      \draw[] (0.5,.865)-- (0.5,2.)  node[at start,right,font=\tiny] {a};
      \draw[] (-0.5,.865)-- (-0.5,2.) node[left,font=\tiny] {c};
      \draw[] (-0.5,2)-- (0.5,2) node[right,font=\tiny] {b};

\end{tikzpicture}
}
\qquad\qquad
\subfloat[Region $F_1$]{\label{fig:B}
\pgfmathsetmacro{\myxlow}{-1}
\pgfmathsetmacro{\myxhigh}{1}
\pgfmathsetmacro{\myiterations}{1}
\begin{tikzpicture}[scale=2]
            \draw[very thin, blue] (0.5,0)-- (0.5,2.2);
              \draw[very thin, blue] (-0.5,0)-- (-0.5,2.2);
    \draw[-latex](\myxlow-0.1,0) -- (\myxhigh+0.2,0);
    \pgfmathsetmacro{\succofmyxlow}{\myxlow+0.5}
    \foreach \x in {\myxlow,\succofmyxlow,...,\myxhigh}
    {   \draw (\x,0) -- (\x,-0.05) node[below,font=\tiny] {\x};
    }
    \foreach \y  in {1}
    {   \draw (0,\y) -- (-0.05,\y) node[above left,font=\tiny] {\pgfmathprintnumber{\y}};
   }
    \draw[-latex](0,-0.1) -- (0,2.4);
    \begin{scope}   
        \clip (\myxlow,0) rectangle (\myxhigh,1.1);
        \foreach \i in {1,...,\myiterations}
        {   \pgfmathsetmacro{\mysecondelement}{\myxlow+1/pow(2,floor(\i/3))}
            \pgfmathsetmacro{\myradius}{pow(1/3,\i-1}
            \foreach \x in {-2,\mysecondelement,...,2}
            {   \draw[very thin, blue] (\x,0) arc(0:180:\myradius);
                \draw[very thin, blue] (\x,0) arc(180:0:\myradius);
            }   
        }
    \end{scope}

    \draw[name path=path1] (-0.5,.865) arc(120:60:1);
        \draw[name path=path2] (-0.5,.865) arc(60:0:1);
              \draw[name path=path3] (0,0) arc(180:120:1);
        \path[name intersections={of=path1 and path2,by=x12}];
        \path[name intersections={of=path3 and path2,by=x23}];
        \path[name intersections={of=path3 and path1,by=x31}];
  \begin{scope}
         \filldraw[draw=gray, opacity=0.,fill=gray,opacity=.5] (-0.5,.865) arc(120:60:1)  -- (0.5,.865) arc(120:180:1)-- (0,0) arc(0:60:1);       
  \end{scope}

     \draw[name path=path1] (-0.5,.865) arc(120:60:1) node[at start,left,font=\tiny] {\(\tilde a\)};
       \draw[name path=path2] (-0.5,.865) arc(60:20:1) node[left,font=\tiny] {\(\tilde b\)};
              \draw[name path=path3] (-0.065,0.34) arc(105:75:.25) node[right,font=\tiny] {\(\tilde c\)};
               \draw[name path=path2] (0.061,0.34) arc(160:120:1) node[right,font=\tiny] {\(\tilde d\)};

 \end{tikzpicture}

}

\caption{Fundamental regions.} 
\label{fund_r}
\end{figure}

As an example we consider the region $F_1$ which is obtained from the fundamental region $F_0$ by applying $S$ transformation, see Fig. \bref{fund_r}.  In fact, one could think that this choice even more preferable, because the upper boundary element of $F_0$ maps to point $\tau=0$, while $(\tilde{d},\tilde{a})$ was considered previously and other two  elements $(\tilde{a},0)$, $(\tilde{d},0)$ are mapped to each other by certain modular transformation (see below), so that similar consideration as in Section~\ref{sec:3}, based on the transformation property of the logarithmic field, can be used to evaluate corresponding contributions explicitly. However, the expression for one-point amplitude has a singularity at $\tau=0$ which requires some care.

Let us illustrate how to come to the formula (\ref{rewrcont3}) for this choice of fundamental domain, starting with (\ref{cont10}).
First of all, the contribution of an integral over the arc $(\tilde{a}, \tilde{d})$ is given by the same formula (\ref{cont3}), but with a different sign:
\begin{equation}
- \frac{1}{4} (2\pi)^2  \frac{1}{2} \frac{\pd}{\pd a}\Delta^\text{L}_a \cdot 2 \int \limits_0^{\pi/6} d\varphi\, \langle O \rangle_{\tau = i \exp (i\varphi)}  \label{cont32}
\end{equation}
since for a new fundamental domain outward normal direction is different. The two arcs $(0,\tilde{a})$ and $(0, \tilde{d})$  can be parametrized respectively as
\begin{equation}
\begin{cases}
\tau_1 = r_1 e^{i\varphi_1}-1,\quad r_1 = 1, \;\; \varphi_1 \in (0, \pi/3) \\
\tau_2 = 1-r_2 e^{-i\varphi_2},\;\; r_2 = 1,\;\;  \varphi_2 \in (0, \pi/3) \\
\end{cases}
\end{equation}
These two arcs are glued together using a modular transformation 
\begin{equation}
\tau_2 = \frac{\tau_1}{\tau_1+1}\;,\quad\text{ or }\quad\varphi_1 = \varphi_2\;,\quad r_2 = \frac{1}{r_1}
\end{equation}
under which the correlation function of primaries transforms as
\begin{equation}
f_\Delta \left(\frac{\tau}{\tau+1} \right) = |\tau+1|^{2\Delta} f_\Delta(\tau)\;.
\end{equation}
Then for the logarithmic operator $O'$ we have
\begin{equation}
\langle O' \rangle_{\tau/(\tau+1)} = \langle O' \rangle_{\tau} + \Delta^{\text{L}'}_a \log |\tau+1|\times  \langle O' \rangle_{\tau_1}\;.
\end{equation}
We integrate over these two arcs the normal derivatives of this correlator (i.e. derivatives with respect to $r_1 = |\tau_1+1|$ and $r_2$, with additional minus sign). Differentiating the transformation rule above with respect to $r_1$, as before, we get that only the term with $\Delta'$ survives when we sum the two contributions; it integrates to 
\begin{equation}
+ \frac{1}{4} (2\pi)^2 \frac{\pi}{3} \Delta^{\text{L}'}_a \langle O \rangle \;.\label{cont4}
\end{equation}
The two terms (\ref{cont32}) and (\ref{cont4}) add to the same answer as in (\ref{cont3}).

Now let us carefully deal with the singularity at $\tau=0$. Let us regularize the boundary adding the arc $\gamma_0 \equiv (\tilde{b}, \tilde{c})$ of small circle of radius $r$ with the center at $\tau=0$.  It is easy to calculate that the length of this arc is $l \approx r^2$ for $r \to 0$. From modular properties we know that 
\begin{equation}
\langle O' \rangle_{i r e^{i\varphi}} = \langle O' \rangle_{i e^{i\varphi}/r} -  \Delta^{\text{L}'}_a \log r \langle O \rangle_\tau\;.
\end{equation}
For $\tau \to i\infty$, $\tau_2 \approx (-i\tau) = \frac{1}{r}$ and $\frac{\pd}{\pd \tau_2} \langle O' \rangle \approx \text{const},\,\tau_2 \to +\infty$, as we discussed in the main part of the work. Because of that, the considered integral over the arc $\gamma_0$ can be rewritten as
\begin{equation}
\frac{1}{4} (2\pi)^2 \int \limits_{\gamma_0}dl\, \left(-\frac{\pd}{\pd r} \langle O' \rangle_{i r e^{i\varphi}} \right) =\frac{1}{4} (2\pi)^2  \int \limits_0^{r^2} dl\, \frac{1}{r^2} \frac{\pd}{\pd \tau_2} \langle O' \rangle_{\tau_2 \to \infty} + \frac{1}{4} (2\pi)^2 \Delta^{\text{L}'}_a\int \limits_0^{r^2} dl\,\frac{1}{r}  \langle O \rangle\;.
\end{equation}
The second term tends to zero when $r \to 0$; thus, we are left with only the first one, which reproduces contribution from infinity (\ref{cont1}) for the original fundamental domain $F_0$.

\subsection{Comments on generalized minimal gravity}
In some contexts it might be interesting to generalize these results to a gravity theory, in which the matter sector has generic central charge $c<1$ (equivalently, non-rational and negative $b^2$) and the structure constants are analogous to (\ref{min3pf}). There are several possible definitions of such ``generalized minimal models'' (GMM). In the original work by Zamolodchikov~\cite{Zamol3pt} this term is used to describe a  theory with continuous set of primary fields $\Phi_\alpha,\,\alpha \in \mathbb{C}$ and the structure constants given by (\ref{min3pf}) with additional identification of fields $\Phi_\alpha$ and $\Phi_{b^{-1}-b-\alpha}$. In a more recent work~\cite{clessthan12015} this notion, GMM, is used for a theory with an irrational central charge, but a discrete set of fields organised as an infinite Kac table, while the model with the continuous spectrum introduced in~\cite{Zamol3pt}
is referred to as ``$c<1$ Liouville theory''.\footnote{We note that the normalization of the primary fields is different in the references~\cite{Zamol3pt} and~\cite{clessthan12015}.}

In \cite{clessthan12015} it is argued that the GMM with the discrete spectrum is ill-defined on higher genus Riemann surfaces: there is a discrete set of possible correlators and all of them are either zero, if they do not satisfy the fusion rules, or infinite, similar to the partition function. On the other hand, in $c<1$ Liouville theory the OPE is always continuous and has no discrete terms, even in the presence of degenerate fields (the analytic structure constant (\ref{min3pf}) does not obey the fusion rules), which makes the matter sector correlator difficult to calculate, even if it turns out to be finite. It would be interesting to see if the answer that we found, being analytic in $p$, could be continued to the generalized case where $p$ is non-integer.

\section*{Acknowledgments}

We thank Alexander Belavin, Alexey Litvinov, Mikhail Lashkevich and Sergey Parkhomenko for useful discussions.

\appendix
\section{Equations of motion on the torus in Liouville CFT} \label{eomliouv}
\paragraph{Regularisation. } EOM in Liouville theory on the torus read
\begin{equation}
\pd \br{\pd} \phi = \pi b e^{2b \phi} \label{eom}
\end{equation}
(we put $\mu = 1$). Average on the torus of the RHS is given by (see e.g. \cite{Belavin:2010sr})
\begin{equation}
\pi b \langle e^{2b \phi} \rangle_\tau = \frac{1}{16 \pi^2} \frac{1}{|\eta(q)|^2 \tau_2^{3/2}} = \frac{1}{16\pi^2 \tau_2^{3/2} (q \br{q})^{1/24}} \left(1 + q +2 q^2 + 3 q^3 + \dots \right) \left(1 + \br{q} +2 \br{q}^2 + 3 \br{q}^3 + \dots \right) \label{2eq}
\end{equation}
which is obtained using eq. (\ref{1pfexlio}) and
\begin{equation}
 C^{Q/2+iP}_{b, Q/2+iP}  = \frac{4P^2}{\pi b}\;. \label{strcon}
\end{equation}
In this appendix we show how to obtain the same result from the LHS of~\eqref{eom}. Defining $\varphi$ as $\frac{1}{2} \frac{\pd}{\pd a} \underbrace{V_a}_{ e^{2a\varphi}} \mid_{a=0}$, we have two problems: 1) the correlator $\langle \pd \br{\pd} \phi \rangle$ seems to vanish due to translation invariance of one-point functions; 2) the structure constant, entering the expression for the one-point function
\begin{equation}
C^{Q/2+iP}_{a, Q/2+iP} = (\pi \mu \gamma(b^2) b^{2-2b^2})^{-a/b} \frac{\Upsilon(b) \Upsilon(2a) \Upsilon(2iP) \Upsilon(-2iP)}{\Upsilon^2(a) \Upsilon(a+2iP) \Upsilon(a-2iP)} \approx \frac{2}{a}\;,\quad a \to 0
\end{equation}
 is singular in the limit $a \to 0$ (even more so if we differentiate with respect to $a$). This divergence appears due to  continuous spectrum of physical states.\footnote{E.g., for free bosonic string genus~1 partition function we also have this divergence, which is related to the target space volume.} To solve these problems, we need to introduce some regularization.  
 The explicit expression of the  correlator
\begin{equation}
  \langle V_a \rangle_\tau = \sum \limits_{\Delta, \lambda, \br{\lambda}} q^{\Delta(\lambda) - \frac{c_L}{24}} \br{q}^{\Delta(\br{\lambda}) - \frac{c_L}{24}} \bra{\Delta, \lambda, \br{\lambda}} V_a(z) \ket{\Delta, \lambda, \br{\lambda}}\;.
\end{equation}
Restricting first the sum to the primary field level, one can rewrite the matrix elements in terms of the correlators on the sphere with complex coordinate $\zeta = e^{z}$ in the following way: 
\begin{equation}
\bra{\Delta_3} V_1(\zeta_1) \ket{\Delta_2} = \lim_{\zeta_3 \to \infty, \zeta_2 \to 0} |\zeta|_3^{4\Delta_3} \langle V_{3}(\zeta_3) V_1 (\zeta_1) V_2(\zeta_2) \rangle = \frac{C(a_1,a_2,a_3)}{|\zeta_1|^{2 (\Delta_1 + \Delta_2 - \Delta_3)}}\;.
\end{equation}
We need to transform the coordinates from sphere to cylinder/torus, which leads to transformation $V_1(z) = V_1(\zeta_1) |\zeta_1|^{2 \Delta_1}$. Then the matrix element on a cylinder
\begin{equation}
\bra{\Delta_3} V_1(z) \ket{\Delta_2} = \frac{C(a_1,a_2,a_3)}{|\zeta_1|^{2 (\Delta_2 - \Delta_3)}}
\end{equation}
is translation invariant (independent of $z$) if $\Delta_2 = \Delta_3$. Only such elements appear in the trace for the torus correlator. 

With these observations, we note that introducing the position dependence as well as regularizing the diagonal structure constant can both be done by replacing the trace with the following expression
\begin{equation}
  \langle V_a \rangle_\tau = \lim \limits_{\epsilon \to 0} \sum \limits_{\Delta, \lambda, \br{\lambda}} q^{\Delta(\lambda) - \frac{c_L}{24}} \br{q}^{\Delta(\br{\lambda}) - \frac{c_L}{24}} \bra{\Delta(P), \lambda, \br{\lambda}} V_a(z) \ket{\Delta(P+\epsilon), \lambda, \br{\lambda}}\;.
\end{equation}
\paragraph{Level zero. } First, let us check what happens at level zero $\lambda = \br{\lambda} = 0$. The matrix element and its derivative are given by 
\begin{align}
&\bra{\Delta(P)} V_1(z) \ket{\Delta(P + \epsilon)} = \frac{C_{a_1,Q/2+ i (P+\epsilon)}^{Q/2 + iP}}{|\zeta_1|^{4 P \epsilon + 2 \epsilon^2}}\;, \nonumber \\
& \pd\br{\pd} \bra{\Delta(P)} V_1(z) \ket{\Delta(P + \epsilon)} = 4P^2\epsilon^2 (1 + O(\epsilon)) \frac{C_{a_1,Q/2+ i (P+\epsilon)}^{Q/2 + iP}}{|\zeta_1|^{4 P \epsilon + 2 \epsilon^2}}\;,
\end{align}
with the structure constant 
\begin{equation}
    C_{a_1,Q/2+ i (P+\epsilon)}^{Q/2 + iP} =  (\pi \mu \gamma(b^2) b^{2-2b^2})^{-(a+i\epsilon)/b}  \frac{\Upsilon(b) \Upsilon(2a) \Upsilon(2iP) \Upsilon(-2i(P+\epsilon))}{\Upsilon(a+i \epsilon) \Upsilon(a-i \epsilon) \Upsilon(a+i \epsilon +2iP)  \Upsilon(a - i\epsilon -2iP)}\;.
\end{equation}
We differentiate it w.r.t. $a$ and take the limit $a \to 0$, keeping in mind that only terms coming from differentiation of $\Upsilon(2a)$ are relevant, otherwise for nonzero $\epsilon$ we get zero in the limit. Taking into account $\Upsilon'(0) = \Upsilon(b)$, we get
\begin{equation}
 \frac{1}{2} \frac{\pd}{\pd a} C_{a_1,Q/2+ i (P+\epsilon)}^{Q/2 + iP} \mid_{a=0} =  (\pi \mu \gamma(b^2) b^{2-2b^2})^{-i\epsilon/b}  \frac{\Upsilon(b)^2 \Upsilon(2iP) \Upsilon(-2i(P+\epsilon))}{\Upsilon(i \epsilon) \Upsilon(-i \epsilon) \Upsilon(i \epsilon +2iP)  \Upsilon(- i\epsilon -2iP)} \approx \frac{1}{\epsilon^2} + \dots
\end{equation}
Now we multiply these results and keep only the finite contribution for $\epsilon \to 0$. We obtain the leading  contribution in $q$ for the LHS of (\ref{eom})
\begin{equation}
\langle \pd \br{\pd} \phi \rangle = \int \frac{dP}{4\pi} \times  4P^2\times (q \br{q})^{-1/24 +P^2} \times (1 + \dots)\;,
\end{equation}
which coincides with the integral for RHS of EOM, taking into account (\ref{strcon}).
\paragraph{Level one. } The matrix elements with descendants can be expressed using the following commutation relations
\begin{equation}
[L_n, V_a(z)] = \zeta^n \left(\pd V_a + n \Delta_a V_a \right)\;.
\end{equation}
For example, on the first level we get
\begin{align*}
& \bra{\Delta} L_1 V_a(z) L_{-1} \ket{\Delta+\delta} = 2 \bra{\Delta} V_a L_0 \ket{\Delta+\delta}+ \zeta (\pd + \Delta_a) \bra{\Delta}  V_a L_{-1} \ket{\Delta+\delta} = \\
&=2 (\Delta + \delta) \bra{\Delta} V_a \ket{\Delta+ \delta}+ \zeta^{-1} (\pd - \Delta_a) \zeta (\pd + \Delta_a) \bra{\Delta}  V_a \ket{\Delta+ \delta} = \\
&= \frac{C_{a_1,a_{\Delta + \delta}}^{a_\Delta}}{|\zeta|^{2\delta}} \left[2 (\Delta + \delta) +(1-\delta - \Delta_a) (\Delta_a - \delta) \right]\;.
\end{align*} 
Note the same $\zeta$-dependence as we had for level zero. Again, this dependence  disappears for $\delta = 0$, as expected from translation invariance property of the torus correlator. This means that additional differentiation $\pd \br{\pd}$ brings out the same factor $\epsilon^2$ which cancels $1/\epsilon^2$ singularity of structure constant's derivative at $a = 0$. The kinematic factor in the brackets is regular and non-zero as $a, \epsilon \to 0$; that means that in our limiting procedure we don't need to differentiate it with respect to $a$ nor leave $\epsilon$-dependence. With appropriate normalization $\bra{\Delta} L_{1} L_{-1} \ket{\Delta} = 2 \Delta$, it gives coefficient $1$ for $O(q)$ contribution to the toric conformal block, as expected from (\ref{2eq}).

In general, at higher levels we do not get any new subtleties and reasoning analogous to the original proof of HEM on the sphere \cite{HEM} (based on the fact that the conformal properties of $\pd \br{\pd} \varphi$ and $e ^{2b\varphi}$ are the same) should work for descendant contributions to the torus correlators.

\bibliographystyle{JHEP}
\bibliography{mlg2}

%\printbibliography

\end{document}